# Femtosecond laser writing of the depressed cladding buried channel waveguides in ZnS crystal

E. Sorokin[1,3], A. Okhrimchuk[1,4], N. Tolstik[2,3], M. Smayev[1], V. Likhov[1], V. L. Kalashnikov[2], and I. T. Sorokina[2,3]

[1]*Mendeleyev University of Chemical Technology of Russia, Miusskaya sq. 9, Moscow, Russia*
[1]*Department of Physics, Norwegian University of Science and Technology, N-7491 Trondheim, Norway*
[2]*Institut für Photonik, TU Wien, Gusshausstrasse 27/387, A-1040 Vienna, Austria*
[3]*ATLA lasers AS, Richard Birkelands vei 2B, 7034 Trondheim Norway*
[4]*Fiber Optics Research Center of RAS, Vavilova str. 38, Moscow, Russia.*

**Abstract:** We report the first direct femtosecond laser-writing of buried channel waveguides in monocrystalline ZnS. We also report the first single-mode Cr:ZnS depressed cladding buried waveguide laser manufactured by femtosecond laser writing. The laser yields 150 mW average power at 2272 nm wavelength with 11% slope efficiency. A depressed cladding waveguide with propagation loss of 0.62 dB/cm at 1030 nm allowed to obtain spectral broadening under femtosecond-pumping at 1030 nm.
**OCIS codes:** Guided waves (130.2790) ; Laser materials processing (140.3390); Infrared and far-infrared lasers (140.3070 ).

## 1. Introduction.

The channel waveguide architecture is promising for the mid-IR supercontinuum generation in media with high Kerr nonlinearity. Recently we reported the first experiments on supercontinuum and frequency comb generation from Cr:ZnSe and Cr:ZnS lasers using various nonlinear fibers such as e.g. germanate and chalcogenide fibers[1,2] as well as chalcogenide nanospikes[3]. The direct femtosecond laser writing paves the way towards unique opportunities opening up upon formation of such waveguides in a non-linear media for supercontinuum generation[4] as well as for Kerr-Lens mode-locking[5]. This technology will make the ultra-short pulsed or supercontinuum source not only extremely compact but also will allow operation at high repetition rates in excess of 10 GHz, which is highly attractive for high resolution and high sensitivity spectroscopy using coherent mid-infrared frequency combs. Further scaling the pulse energy and peak power will also open up the way towards highly efficient compact few optical cycle frequency comb sources around 5 microns – the work similar to the one previously done with bulk Cr:ZnSe[5] or Yb- and Tm-fiber lasers[6]. In this paper we report on the first steps towards creation of a single mode channel waveguide in highly non-linear ZnS single crystal and using it for supercontinuum generation.

## 2. Direct laser writing of a waveguides in ZnS and ZnS:Cr crystals.

### 2.1 Experimental

The experimental setup is shown in Fig. 1

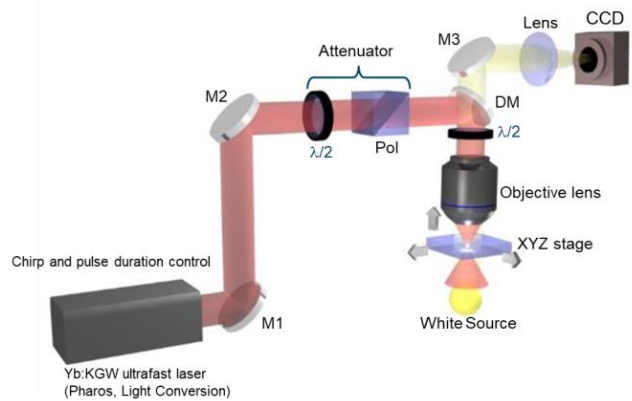

Figure 1. Experimental setup for a waveguide writing.

For direct laser writing of the waveguide we used a femtosecond Yb:KGW based laser system with a regenerative amplifier operating at the wavelength $\lambda = 1028$ nm at 100 kHz repetition rate. We focused the laser beam perpendicular to the sample scanning direction. The sample was mounted on a 3-D precision translation stage with an air bearing from Aerotech. An Olympus objective with NA=0.85 and the spherical aberration correction was used for focusing the laser beam under the polished crystalline surface. We used the rather chirped femtosecond pulses with duration of 800 fs (FWHM) and energy of 350 nJ. We choose the pulse duration of 800 fs as a compromise between the high refractive index change and smoothness and homogeneity of the track. The pulse energy was as high as 350 nJ, which corresponds to a plateau in the plot of the refractive index change on the pulse energy (Fig.2)

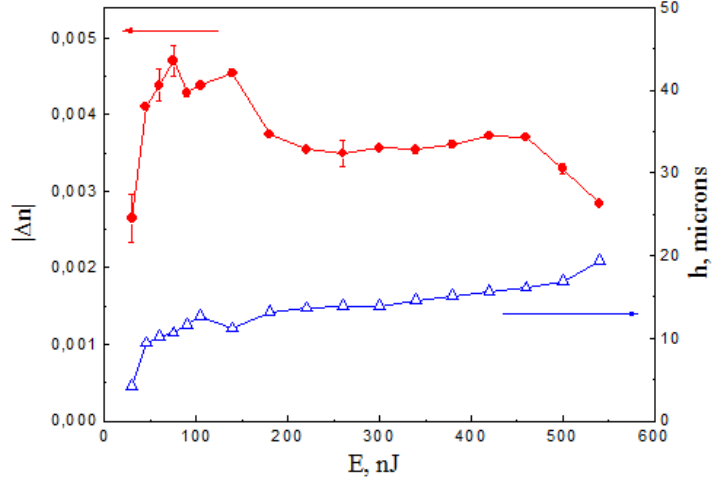

Figure 2. Dependence of refractive index change $\Delta n$ and track height $h$ on the pulse energy $E_p$ ($\tau = 800$ fs, $f = 100$ kHz, $V = 0.25$ mm/s).

By varying the above laser parameters and the waveguide core parameters, we could produce five different depressed cladding waveguides consisting of parallel tracks of the laser modified regions - defects. The waveguides varied by the number of tracks and a core diameter. We kept the distance between the tracks to be 3 μm in all waveguides, and the core diameter was controlled through the number of tracks. The front view microscope picture of the waveguide consisting of 42 tracks is shown in Fig. 3.

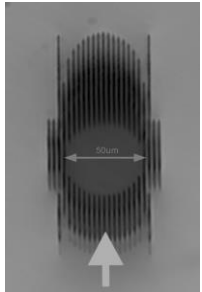

Figure 3. End view of the waveguide written in ZnS:Cr2+ crystal. The arrow indicates direction the laser writing beam.

### 2.1.1 Optimization of writing parameters.

A Yb:KGW femtosecond laser system with a regenerative amplifier operating at wavelength $\lambda = 1028$ nm was used for direct laser writing in standard setup, where the focused laser beam is perpendicular to the direction of the sample scanning. An objective length with NA=0.85 was used for focusing at depth of 100 μm under the polished crystal surface. Sample was held on the 3-D precision translation stage with air bearing (Aerotech). Trial writing of tracks was done with different pulse duration, pulse repetition rates and scatting velocity of the sample. Refractive index change in the tracks was investigated with QPm technique. At first dependence of refractive index change upon pulse duration was obtained with repetition rate $f$ and scanning velocity $V$ fixed at 5 kHz and 0.25 mm/s correspondingly (Fig.1).

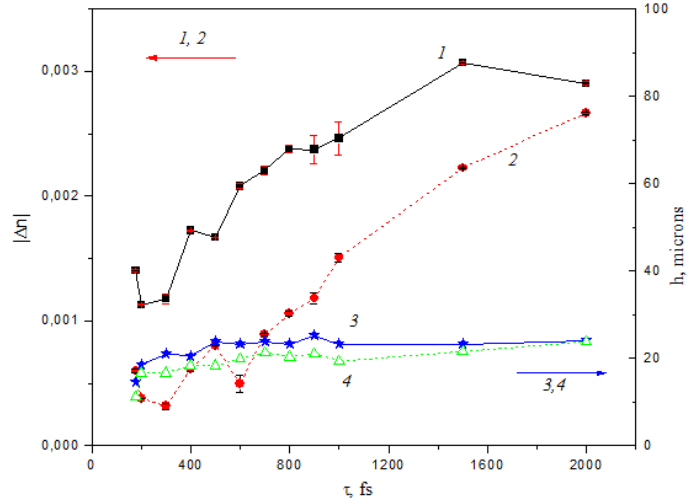

**Figure 3**. Dependencies of the refractive index change $|\Delta n|$ (1,2) and the track height $h$ (3,4) on the pulse duration $\tau$ ($f = 5$ kHz, $E_p = 960$ nJ, $V = 0.25$ mm/s). *1, 3* – Polarization $E$ is perpendicular to the scanning direction $s$, *2, 4* – $E$ is parallel to $s$.

All obtained tracks have the decreased refractive index with respect to non-modified regions of the crystal. The tracks were fairly smooth and homogeneous under pulse duration not exceeding 1000 fs, and visible inhomogeneity was observed in microscope as pulse duration increased over 1000 fs. The refractive index change grows significantly with pulse duration while the rack height remains almost unchanged and equal to $h \approx 20$ μm. Refractive index change $|\Delta n|$ produced by the beam with polarization perpendicular to the scanning direction is larger than $|\Delta n|$ produced by the beam with polarization parallel to the scanning direction. Pulse duration of 800 fs was chosen as a compromise between high refractive index change and homogeneity of a track in experiments for further optimization of writing on the pulse energy, repetition rate and scanning velocity. The tracks were slightly rugged at high repetition rates and writing velocities. So, the intermediate conditions ($f = 180$ kHz; $V = 5$ mm/s) were selected for waveguide writing in order to avoid any inhomogeneity in tracks.

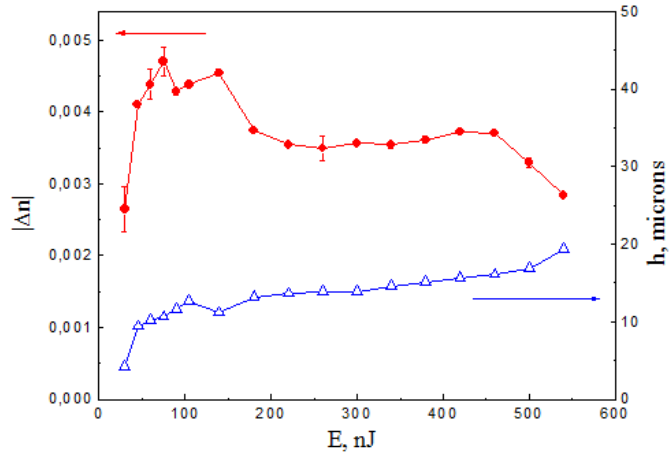

**Figure 4.** Dependencies of refractive index change $|\Delta n|$ and track height $h$ on the pulse energy $E_p$ ($\tau = 800$ fs, $f = 5$ kHz, $V = 0.25$ mm/s, polarization was perpendicular to the scanning direction).

Fig. 4 shows the dependence of refractive index change $|\Delta n|$ obtained at $\tau = 800$ fs on the pulse energy $E_p$. The polarization of light beam was perpendicular to scanning direction. The dependence is not monotonic: fast initial growth followed by slow decrease of the refractive index change. The highest value of refractive index change was found to be $|\Delta n|_{max} = (4.7 \pm 0.2) \cdot 10^{-3}$.

Then we inscribed tracks under opposite directions of the sample movement (s and -s) in order to reveal possible quill wring. Significant difference in refractive index change for these tracks was not found (Fig. 5), that manifests that the quill writing is not sufficient for our laser beam[7].

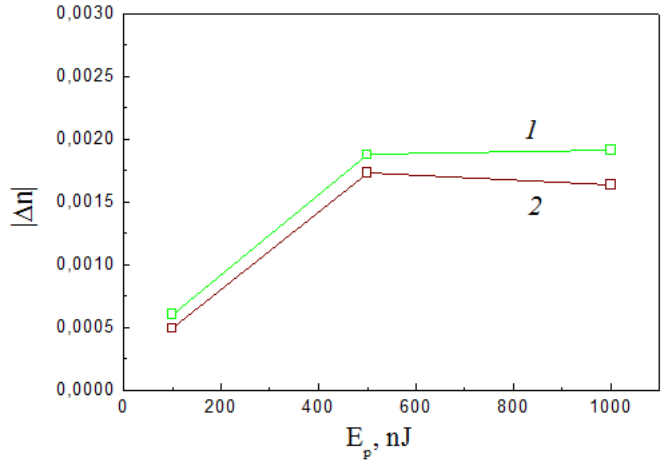

**Figure 5.** Dependencies of refractive index change |Δn| on the pulse energy $E_p$ at opposite directions of laser scanning 1 – positive *s*, 2 – negative *s* ($\tau$ = 178 fs, $f$ = 5 kHz, $V$ = 0.25 mm/s, polarization was perpendicular to the scanning direction).

Optimum writing conditions include suitable energy range, pulse duration, repetition rate and velocity of sample movement. In order to optimize the last two parameters the laser writing was investigated under repetition rate *f* and velocity of sample movement *V* in the ranges of 1 ÷ 800 kHz and 0.1 – 30 mm/s, correspondingly, while keeping constant number of laser pulses per unit length of a track, that is *f/V*=const. Dependences of the refractive index change |*Δn*| on scanning velocity *V* at various repetition rate *f* are presented at Fig.6. The tracks were slightly rugged at high repetition rates and writing velocities. So, the intermediate conditions (*f* = 180 kHz; *V* = 5 mm/s) were selected for waveguide writing in order to avoid any inhomogeneity in the tracks.

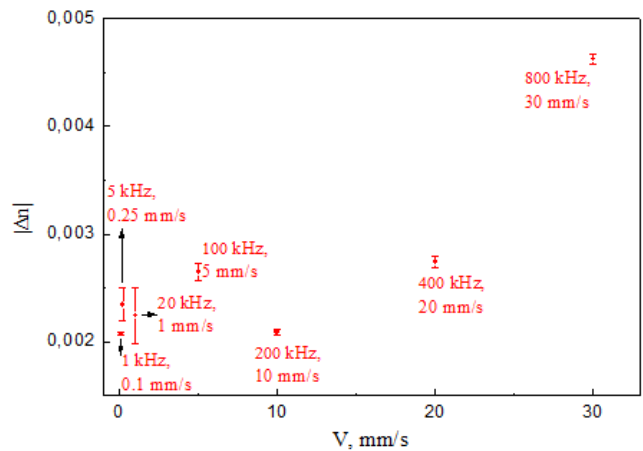

**Figures 6.** Dependence of refractive index change |*Δn*| on the velocity *V* of light beam movement at various repetition rates *f*, polarization was set perpendicular to the scanning direction, $\tau$ = 800 fs, Ep = 500 nJ.

### 2.3. Depressed cladding waveguide writing.

We used depressed cladding architecture with a tube-like cladding for channel waveguide writing [8,9]. Repetition rate, scanning velocity, pulse duration and pulse energy were set to 180 kHz, 5 mm/s, 800 fs and 500 nJ correspondingly. Non-linear dependence of the track depth upon z-coordinate position was taken into account in the programming code for Aerotech translation stage. Series of waveguides differed by number of tracks in the cladding and distance between them were written in ZnS and ZnS:$Cr^{2+}$ crystals. Waveguide inscribed in ZnS possessing lower number of transverse modes at 1030 nm are shown in Fig.7.

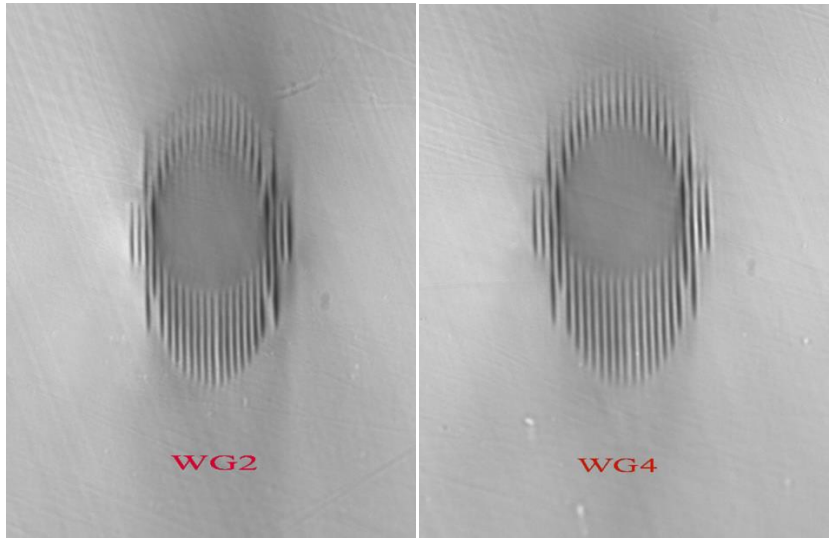

**Figure 7.** Waveguide structures WG2 and WG4 inscribed in ZnS crystal by femtosecond laser pulses ($\lambda$ = 1028 nm, $\tau$ = 800 fs, f = 100 kHz, V = 5 mm/s, Ep = 500 nJ).

## 3. Characterization of the depressed cladding waveguide.

The beam of Yb:KGW femtosecond laser system TETA (Avesta Ltd., $\lambda$ = 1028 nm, repetition rate 100 kHz, pulse duration 300 fs) was coupled to a waveguide by a 2-lense system (Fig. 8). Beam diameter at the waveguide input and at the beam waist was adjusted by changing distance between the lenses, and was nearly matched to fundamental mode radius. Intensity distribution at the output of inscribed waveguides was imagined with a lens on the Spiricon SP620U sensor (Ophir) connected to PC and controlled by BeamGage Standard 6.5 software. For analysis of spectral properties of radiation at the waveguide length optical spectrum analyzer AQ6370D (Yokogawa) was utilized.

The propagation loss was determined by direct measurement of the waveguide transmittance: the power passing through the waveguide and the collecting lens was divided by the power passing through the collecting lens without the sample, when this lens collected the light directly from the input lens (the collecting lens was shifted to the input lens by distance equaled to waveguide length). The measured total loss was the same for both waveguides at the level of 1.58 dB/cm. In the process of calculations Fresnel reflectance losses were taken into account. Large total loss may be explained by large absorbance of the sample under study at $\lambda$ = 1028 nm. Measured value of absorption loss was 0.96 dB/cm. Thus, waveguide loss for inscribed structures was as low as 0.62 dB/cm.

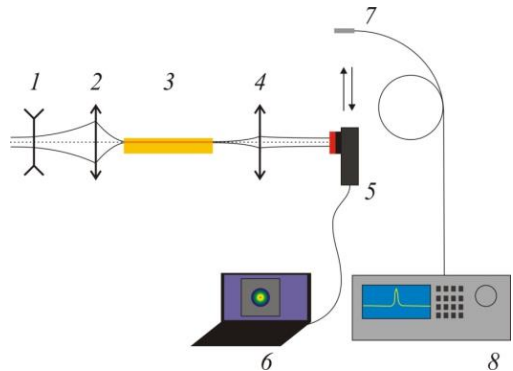

**Figure 8.** Scheme of experimental setup for waveguide investigation: 1, 2, – plane- concave and plane- convex lenses correspondingly (f1 = - 500 mm, f2 = 150 mm); 3 – ZnS crystal sample; 4 - 10x objective lens, NA = 0.25, 5 – beam profile analyzer; 6 – personal computer; 7 – optical fiber; 8 – optical spectrum analyzer.

It was found that both inscribed structures shown in Fig.9 number of modes is increased, when average power exceeds nearly 1 mW. Correspondingly we have found a broadening of the spectrum at the waveguide output at this level of average power. The increase of the beam power results in growth of spectral width up to 70 nm, which was obviously restricted because of strong positive GVD in the spectral region of investigations.

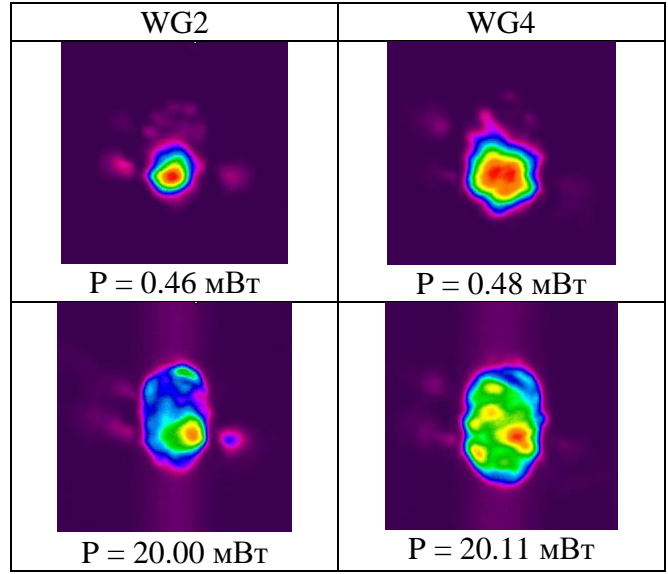

**Figure 9.** Experimental distribution of intensity at the waveguide output for WG2 and WG4 structures at different light power at waveguide input.

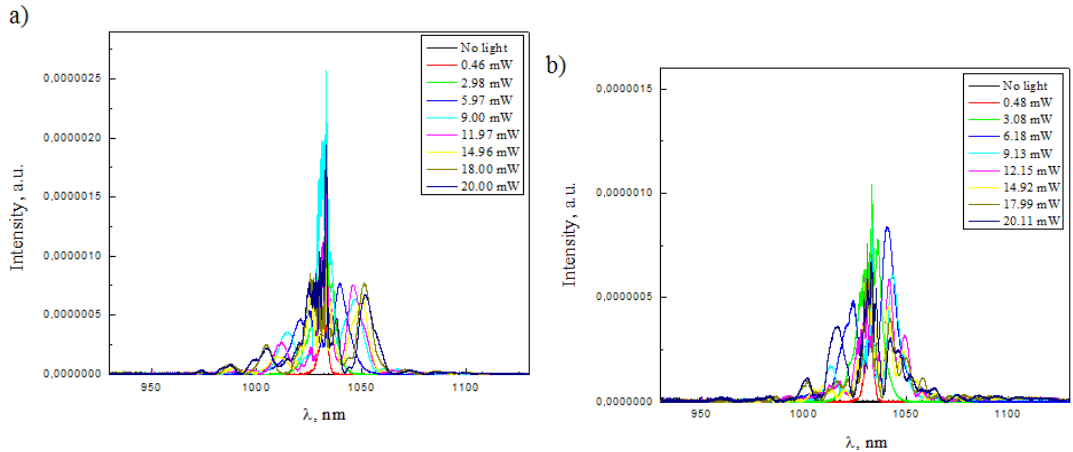

**Figure 10.** Spectra of the output waveguide mode (input beam: λ = 1028 nm, τ = 300 fs, f = 100 kHz) at different beam powers. a) WG2 and b) WG4.

## 4. Single-mode vs. multi-mode wave-guide laser operation

The laser cavity consisted of cavity mirrors butt-coupled to carefully polished facets of the uncoated crystal. The active waveguide was pumped at 1610 nm by the emission of 5W erbium fiber laser focused to the waveguide facet by the plano-convex lens with focal length of 40 mm. The pump power was limited at around 2 W due to the back-reflection from the uncoated crystal facet to the pump laser. Cavity input mirror was made high-reflective (HR) in the range 2100-2500 nm, and transmitted around 95 % of the pump emission at 1610 nm. Output coupler mirror (OC) had transmittance around 10% at laser wavelength. The emission from the waveguide containing both laser emission, and the residual pump light was collimated by 20 mm focal distance lens. Two dichroic mirrors similar to the one used as the cavity input mirror were implemented to filter out the residual pump.

The quality of facet polishing and orthogonality of facets and waveguide direction are critical for successful waveguide laser operation. After the proper facet quality has been reached, the majority of the manufactured waveguides exhibited lasing. Typical output characteristics and laser spectrum are shown on the Fig. 11a. For the incident pump power, the Fresnel reflection and transmittance of the dichroic mirror were accounted for, though the unabsorbed residual pump was not. Slope efficiency of 11% was measured, which is comparable to the performance of waveguide Cr:ZnS laser reported previously. Average output power of 150 mW was obtained which is the highest output power demonstrated so far from a Cr:ZnS waveguide laser. The emission was centered at 2272 nm ( Fig. 11b).

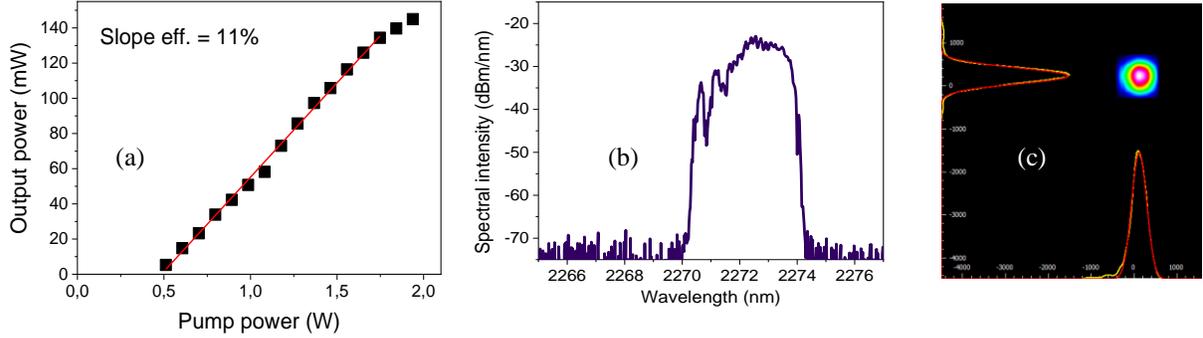

**Figure 11**. Output power (a), emission spectrum (b), and beam shape (c) of a Cr:ZnS waveguide laser.

We have also measured the beam quality from a waveguide laser in a far field using the moving-slit beam profiler. The observed profile was close to Gaussian (Fig. 11c).

We believe that the possibility to create both, single mode as well as multi-mode behavior in the waveguide is largely explained by the complex dynamics in this particular medium, enhanced by the large nonlinearity in ZnS. We analyze this in the next two sections and discuss the opportunities that it opens for both, passive distributed mode-locking as well as supercontinuum generation.

## 5. Modeling of a laser beam dynamics

We based our numerical modeling on a paraxial approximation and the assumption of axial symmetry. The last allows representing the laser ($l$) and pump ($p$) beam intensity in the form of $I_{l,p} = I^0_{l,p}(z)\mathcal{F}(z,r)$, where $I^0_{l,p}(z)$ is a laser/pump intensity on a beam axis ($r = 0$), $z$ and $r$ are the intra-waveguide propagation and radial coordinates, respectively, and $\mathcal{F}(z,r)$ ($\lim_{r\to\infty}\mathcal{F}(z,r) = 0$) defines a laser/pump beam evolution along the waveguide. The following sections can describe the closed intra-waveguide evolution:

1) *Forward propagation* of a complex laser/pump field amplitudes $a^m_{l,p}(z)$ ($|a^m_{l,p}(z)|^2 \equiv I^m_{l,p}$) on the $m^{th}$ cavity round trip

$$\begin{pmatrix} \frac{\partial}{\partial z} a^m_l(z,r) \\ \frac{\partial}{\partial z} a^m_p(z,r) \end{pmatrix} = \begin{pmatrix} \frac{i}{2\beta_l}\Delta_r - \frac{i\,n(r)\,k^0_l}{2}\left(\frac{n(r)}{n_{cl}} - 1\right) + \frac{G(z,r)}{2} \\ \frac{i}{2\beta_p}\Delta_r - \frac{i\,n'(r)\,k^0_l}{2}\left(\frac{n'(r)}{n'_{cl}} - 1\right) \end{pmatrix} \begin{pmatrix} a^m_l(z,r) \\ a^m_p(z,r) \end{pmatrix}, \quad (1)$$

$$G(z,r) = G_1\,G_2\,|a^m_p(z,r)|^2 / \left(1 + G_2\,|a^m_p(z,r)|^2 + G_3\,|a^m_l(z,r)|^2\right). \quad (2)$$

2) Reflection from output mirror

$$b_l^m(z,r) = \sqrt{1-T_l}\, a_l^m(z,r),$$
$$b_p^m(z,r) = \sqrt{R_p}\, a_p^m(z,r). \qquad (3)$$

3) Back-propagation

$$\begin{pmatrix} \frac{\partial}{\partial z} b_l^m(z,r) \\ \frac{\partial}{\partial z} b_p^m(z,r) \end{pmatrix} = \begin{pmatrix} \frac{i}{2\beta_l}\Delta_r - \frac{i\,n_{cl}\,k_l^0}{2}\left(\frac{n(r)}{n_{cl}}-1\right) + \frac{G(z,r)}{2} \\ \frac{i}{2\beta_p}\Delta_r - \frac{i\,n'_{cl}(r)\,k_l^0}{2}\left(\frac{n'(r)}{n'_{cl}}-1\right) \end{pmatrix} \begin{pmatrix} b_l^m(z,r) \\ b_p^m(z,r) \end{pmatrix}. \qquad (4)$$

$$G(z,r) = G_1\, G_2\, |b_p^m(z,r)|^2 \big/ \left(1 + G_2\, |b_p^m(z,r)|^2 + G_3\, |b_l^m(z,r)|^2 \right). \qquad (5)$$

4) Closing on input mirror

$$a_l^{m+1}(z,r) = \sqrt{R_l}\, b_l^m(z,r),$$
$$a_p^{m+1}(z,r) = \sqrt{1-T_p}\, b_p^{m+1}(z,r) + \sqrt{T_p}\, a_p^{in}(r). \qquad (6)$$

Eqs. (1,4) describe the evolution of the axially-symmetrical laser/pump beams $a_l(z,r), a_p(z,r)$ along a waveguide longitudinal coordinate $z$ under the action of diffraction ($\Delta_r = \frac{1}{r}\partial/\partial r(r\,\partial/\partial r)$ is a two-dimensional Laplacian in cylindrical coordinates) and waveguiding supported by an $r$-dependent difference between core/cladding indexes of refraction for a laser field [16]: $n(r) = n_{cl} + (n_{co} - n_{cl})\exp[-r^s/w_0^s]$ at the lasing wavelength $\lambda_l$ =2.272 mm ($n_{co}$ =2.2629, $n_{cl}$ =2.594, $w_0$ is a waveguide radius, and $s$=2,10). $k_l^0 = 2\pi/\lambda_l$ is a laser field wavenumber.

The Gaussian-shaped input pump beam $a_p^{in}(r) = \sqrt{2 T_p P_p^0/\pi w^2}\, \exp\left(-\frac{r^2}{w^2} + i(\psi - k_p^0\, n'(0) r^2/2 R_c)\right)$ with the radius and wavenumber of $w$ and $k_p^0 = 2\pi/\lambda_p$, respectively (the pump transmission coefficient $T_p$ is 0.95, the pump wavelength is of $\lambda_p$=1.61 mm, the input Gouy phase $\psi$=0, and $R_c \gg w_0$) provides amplification and soft-aperture-like action on the laser field through the $z$- and $r$-dependent gain coefficient $G(z,r)$. The pump beam obeys diffraction as well as waveguiding due to profiled refractive index $n'(r) = n'_{cl} + (n'_{co} - n'_{cl})\exp[-r^s/w^s]$ ($n'_{co}$ =2.2706, $n_{cl}$ =2.2671). Its input radius equals the waveguide one in our calculations, i.e., $w = w_0$. $P_p^0$ is a pump power and $T_p$ =0.95 is an input mirror transmission at $\lambda_p$.

The saturable pump evolves in line with Eqs. (2,5) where the saturation effect is integral over a laser round-trip period $T_c$ due to its smallness in comparison with a gain relaxation time $T_r$. The coefficient in Eqs. (2,5) are [17]:

$$G_1 = \sigma_{em} N_{Cr} L_c, \quad G_2 = \frac{\sigma_{abs} T_c}{2\,E_p}, \quad G_3 = \frac{\sigma_{em} T_c}{2\,E_l}, \qquad (7)$$

where $\sigma_{em}$ =1.49× $10^{-18}$ cm² is the active medium emission cross-section, $\sigma_{abs} = 10^{-18}$ cm² is the absorption cross-section, $N_{Cr} = 0.14 \times 10^{18}$ cm⁻³ is the concentration of Cr-ions, $E_p$ and $E_l$ are the energy of pump and laser photons, respectively. The maximal small signal gain for one cavity round-trip is of $4 G_1 G_2 T_p P_p^0/\pi w^2$ which varies from 0.2 to 0.28 in our case ($P_p^0$=1, 2 W and $w$=20, 25 mm).

Eq. (3) describes an action of the output mirror with the transmission $T_l$=0.1 for a laser field and the reflection $R_p$=0.27 for a pump.

Finally, Eq. (6) closes the round-trip (we assume the reflectivity of input mirror $R_l$=1 at $\lambda_p$). The numerical procedure was based on the explicit finite-difference scheme with the initially Gaussian laser beam coinciding with the input pump one and $a_l^1(0,0) = 10^{-3} a_p^1(0,0)$. The results are shortly presented below.

## 6. Numerical results and discussion

The steady-state laser beam and gain profiles obtained after 1000 cavity round-trips are shown in Figs. 12, 13. For a comparatively thin waveguide and pump beam ($w = 20$ μm), the main part of the energy is localized in a fundamental mode (compare solid and dotted curves in Fig. 1). The $M^2$-factor for a smaller pump power (solid black curve) tends to unity. Still, the existence of a «tightened wing» testifies to the contribution of higher-order spatial modes whose profiles are represented by insets (a-d) in Fig. 13.

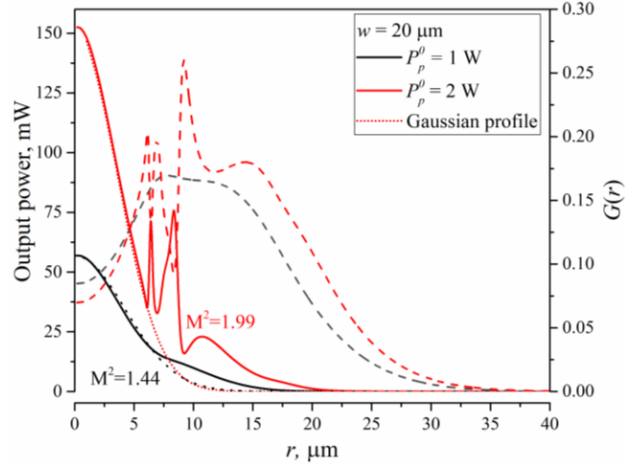

**Figure 12**. Radial profiles of output laser power (solid curves) and dimensionless gain coefficient $G$ (dashed curves) for different pump powers (1 W, black and 2 W, red). Dotted curves show the Gaussian fitting. Pump beam size $w = 20$ μm, $s = 2$.

Such a contribution can be estimated by the expansion of the calculated power profile over the Gauss-Hermitian modal basis that allows calculating a relative weight of modes $C_n$:

$$C_n = \frac{\int_0^R P(r')\, H(n,r')^2 \exp(-2\, r'^2)dr'}{\left(\int_0^R H(n,r')^2 \exp(-2\, r'^2)dr'\right)^2}, \qquad (8)$$

where $P(r')$ is output power, $r'$ is a radial coordinate normalized to a size of a «fundamental mode» (see pointed curves in Fig. 12), and $H(n,r')$ is the Hermite polynomial of $n^{th}$-order ($n = 0,1,2,...$ is a mode order and $R \gg 1$). In this normalization, the beam quality factor $M^2 = \sum_n C_n$. The degeneration of higher-order modes has to be taken into account in Eq. (8). Five lowest modes are sufficient for «covering» an accurate beam profile in our case (see Table I).

Table I. The relative weighting coefficients for transverse modes of Fig. 1.

| Pump power | $C_0$ | $C_1$ | $C_2$ | $C_3$ | $C_5$ |
|---|---|---|---|---|---|
| 1 W | 1 | 2 ×0.175 | 2 ×0.045 | 0.0017 | 0.00017 |
| 2 W | 1 | 2×0.38 | 2×0.105 | 0.016 | 0.002 |

One can see that the pump power growth excites higher-order modes and leads to the $M^2$-growth (red curve in Fig. 12). Nevertheless, an experimentally measured beam profile can mimic the Gaussian one.

Two facts attract attention: hole-burning in the transverse gain profile (dashed curves) and a squeezing of the generation beam compared with the pump one (compare solid and dashed curves in Figure). The first phenomenon results from the gain saturation. The second one will be clarified below.

The growth of a waveguide (and pump beam) size excites the higher-order transverse modes and increases $M^2$ (Fig. 13 and Table II). These effects are lower for a waveguide with a sharper gradient between core and cladding refractive indexes (blue curves in Fig. 13 corresponding to $s=10$).

Table II. The relative weighting coefficients (without degeneration factors) for transverse modes of Fig. 2.

| Pump power | $C_0$ | $C_1$ | $C_2$ | $C_3$ | $C_5$ |
|---|---|---|---|---|---|
| 1 W | 1 | 0.35 | 0.09 | 0.015 | 0.0017 |
| 2 W | 1 | 1.02 | 0.38 | 0.062 | 0.0054 |
| 2 W, $s = 10$ | 1 | 0.43 | 0.115 | 0.018 | 0.0021 |

To adequately represent the mode structure, we must consider that the gain coefficient is radially graded due to a

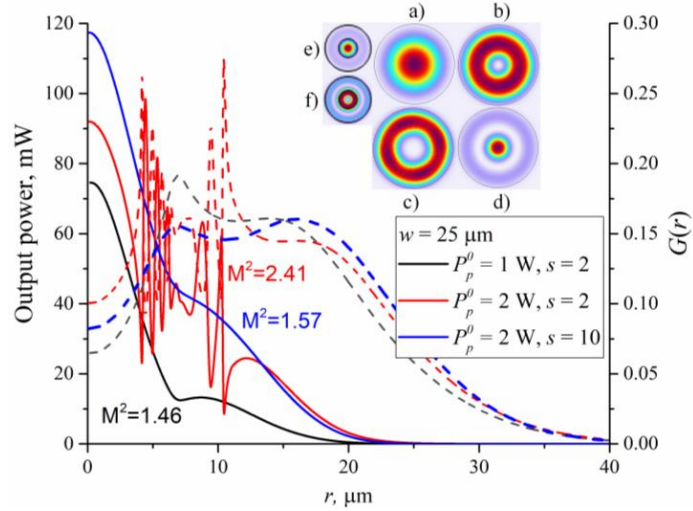

**Figure 14**. Radial profiles of output laser power (solid curves) and dimensionless gain coefficient $G$ (dashed curves) for different pump powers (1 W, black and 2 W, red). Blue curves correspond to the super-Gaussian pump beam with $s = 10$; $w = 25$ μm. Insets (a - d) show profiles of four lowest Hermite-Gaussian modes in Eq. (8). (e-f) correspond to two lowest non-Hermitian modes of a core combined from inner ($w_{in}$=10 μm, $n_{in}=n_{co} - i\, 3.8 \times 10^{-4}$) and outer ($w = 25$ μm, $n(r)$) parts.

Gaussian-shaped pumping beam. As a result, the modes become non-Hermitian. Such a situation, which corresponds to our numerical model (1-6), could be modeled analytically by the insertion of a sub-core $w_{in}$ with a complex index of refraction $n_{in}$ inside a core $w$ (see subsets (e,f) in Fig. 14). This results in an additional mode squeezing or *dissipative mode-cleaning*, under soft-aperture-like action of graded gain. Potentially, this effect could realize the distributed Kerr-lens mode-locking in a waveguide laser predicted in [18] and using a Kerr-lensing of laser beam inside a waveguide.

## 7. Conclusion

Direct laser writing was shown to be a promising technique for manufacturing of low loss single mode channel depressed cladding waveguides in ZnS single crystal. It opens up unlimited opportunities to fabricate both passive as well as active ion doped ZnS/ZnSe based waveguides for integrated photonic devices.

**Acknowledgements**
E.Sorokin, N. Tolstik, V. L. Kalashnikov and I. T. Sorokina would like to acknowledge funding from the Norwegian Research Council projects #303347 (UNLOCK) and #326503 (MIR) as well as FWF project #P24916, and A. Okhrimchuk, M. Smayev, V. Likhov would like to acknowledge funding from Russian Science Foundation project #18-19-00733.